\documentstyle[aps,multicol,epsfig,prl]{revtex}

\input epsf

\def\prb{Phys.\ Rev.\ B}

\def\prl{Phys.\ Rev.\ Lett.\/}

\def\be{\begin{equation}}
\def\ee{\end{equation}}
\def\ba{\begin{eqnarray}}
\def\ea{\end{eqnarray}}

\def\YBCO{YBa$_2$Cu$_3$O$_{7-\delta}$}
\def\248{Y$_2$Ba$_4$Cu$_8$O$_{7-\delta}$}

\def\C60{A$_x$C$_{60}$}

\begin{document}

\title
{Nodal quasi-particles in stripe ordered superconductors}

\author{M.~Granath$^1$, V.~Oganesyan$^1$, S.~A.~Kivelson$^1$, 
E.~Fradkin$^{2}$, and V.~J.~Emery$^3$}
\address
{ Dept.\ of Physics, U.C.L.A.$^1$, Los Angeles, CA  90095;
  Dept.\ of Physics, University of Illinois$^2$, Urbana, IL 61801;
  Dept.\ of Physics, Brookhaven National Laboratory$^3$, Upton, NY 11973}

\date{\today}
\maketitle
\begin{abstract}
We study the properties of a quasi-one dimensional superconductor
which consists of an alternating array of two inequivalent chains.
This model is a simple charicature of a striped
high temperature superconductor, and is more generally  a theoretically
controllable system in which the superconducting state emerges from a
non-Fermi liquid normal state.  Even in this limit, ``d-wave like'' order
parameter symmetry is natural, but the superconducting state can
either have a complete gap in the quasi-particle spectrum,
or  gapless ``nodal'' quasiparticles.  We also find circumstances in
which antiferromagnetic order (typically incommensurate) coexists with
superconductivity.
\end{abstract}

\begin{multicols}{2}
\narrowtext

A key feature of the cuprate high temperature superconductors is
that the ``normal'' state
is not well described as a Fermi liquid.  Therefore, to understand the
physics of the transition temperature, the
brilliantly successful BCS theory, which presupposes\cite{schrieffer}
that the normal
state is a Fermi liquid, must be modified.
Although the cuprate high temperature superconductors are
layered materials,  self-organized one dimensional
structures\cite{PNAS}, or ``stripes,'' have been widely observed,
making plausible the idea that at intermediate
scales these materials can be thought of as quasi-one dimensional.  
The only theoretically
well understood example of a superconducting system with a non
Fermi-liquid (NFL) normal state is a quasi-one dimensional superconductor.
Here the ``normal state''
is governed by the quantum critical physics of a
decoupled set of one dimensional electron gases (1DEG);
superconducting long-range order is triggered by inter-chain coupling,
and accompanied by a crossover to higher dimensional
physics\cite{carlson}. 

Several salient features of the high temperature superconductors are
naturally understood from this viewpoint\cite{carlson}.  The
fact\cite{fedorov} that in
under and optimally doped materials,
quasi-particles in the gap antinodal regions of the Brillouin Zone (BZ)
exist only
in the superconducting state, and have a quasi-particle weight
($Z(T,x)$) which vanishes as the transition point is approached, either
as a function of the doped hole concentration, $x$, or
of the temperature, $T$, can, in our opinion, only be understood if
the normal state has no well defined quasiparticle excitions.  The
extraordinarily one dimensional dispersion apparent\cite{fs} in the
quasi-particle spectrum in this region of the BZ is independent
evidence of a quasi-1d origin for the NFL behavior.
That the zero temperature superfluid density, and with it
$T_{c}$, is roughly proportional to $x$ is also simply understood in
this way, as is the fact that the pairing scale, $\Delta_{0}$, and
the superconducting $T_{c}$ appear to be distinct energy scales in the
problem.

We thus propose studying the high temperature superconductors by adiabatic continuity from the
quasi-one-dimensional limit.  Conceptually, we imagine introducing an explicit symmetry breaking
field of strength $h$ into the physical Hamiltonian, so that for $h$ large the problem is
iterally quasi-one-dimensional, and can hence be {\it solved} using the powerful
non-perturbative methods developed for the theory of the 1DEG.  Then, so long
as there is no
  
\begin{figure}
\begin{center}
\noindent
\epsfxsize=2.5in
\epsfysize=2.5in
\epsfbox{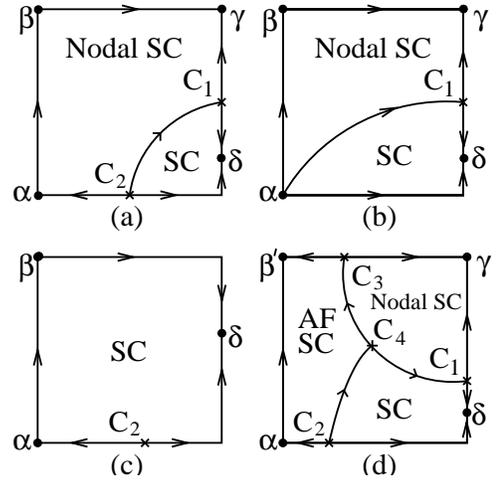}
\end{center}
\caption
{Schematic phase diagrams and qualitative RG flows of 
an array
of alternating A and B type chains with $\Delta_s^{A} >0$ and
$\Delta_c^{A}=\Delta_{c}^{B}=
\Delta_{s}^{B}=0$. The figures represent cuts through a multidimensional
parameter space. Interactions 
that couple neighboring $A$
and $B$ chains increase along the $x$ axis,
while the 
$A$ to $A$  and $B$  to $B$ couplings increase
along $y$. The RG flows in the neighborhood of the fixed points 
(excpect $C_4$) follow from the analysis presented in the text;  the
phase boundaries 
and global flows 
are qualitative renderings.  
The unstable fixed points represent
decoupled chains ($\alpha$), the two fluid
state with a 2D gapped  superconductor on A and
an anisotropic Fermi liquid on B ($\beta$), and various critical points ($C_j$).
The phases controlled by the stable fixed points  are: coexisting
superconducting and antiferromagnetic order with a full quasi-particle gap ($\beta^{\prime}$),
a 2D nodal superconductor ($\gamma$), and  a 2D fully gapped
superconductor ($\delta$). 
}
\label{fig:four}
\end{figure}
\noindent

\noindent{phase transition }
as a function of $h$, the results should be qualitatively correct,
even as
$h\rightarrow 0$.  Moreveover, so long as the isotropic system has substantial local stripe
order, many of these results should even be quantitatively reasonable.   Empirical evidence 
which suggests that such explicit symmetry breaking is innocuous comes from experiments in the
strongly orthorhombic materials, {\YBCO} and {\248}, where anisotropies in the in-plane
superfluid density as large as a factor of 10 can be induced without, apparently, affecting the
qualitative physics of high temperature superconductivity. 

However, there is also experimental evidence of gapless ``nodal''
excitations at low temperatures, deep in the superconducting state.
In all the simplest
realizations of a quasi-one dimensional superconductor, for example
an array of weakly interconnected doped two-leg $t-J$ or Hubbard ladders, the
superconducting state has a gap to all spin-carrying
excitations, including quasi-particles.  
The gapless and fully
gapped phases of a superconductor
are necessarily separated by a quantum phase transition!
In order to substantiate
the claim that the quasi-one-dimensional limit is adiabatically connected to
the physics of the
cuprates,
it is at least {\it necessary} to determine whether, and under what
circumstances,
gapless nodal excitations exist.

It is important to stress that
the issue of the ``d-wave-like'' character of the order parameter
({\it i.e.} whether the expectation value of the pair creation operator
changes sign under a $90^{o}$ rotation)  is distinct from the issue of
the existence of nodal quasi-particles.  Even in a weak coupling (BCS)
superconductor, if the Fermi surface does not close around the $\vec
k=(0,0)$ or $\vec k=(\pi,\pi)$ points in the BZ, it is possible to
have perfect d-wave symmetry without nodal quasi-particles.  For the
two-leg ladder of the above cited example, the superconducting pairing is
known\cite{2leg} to be d-wave-like but fully gapped.

We will study the  simple model
effective Hamiltonian,
$H=H^{*}+H^{\prime}$,
of a quasi-one
dimensional system which consists of an array of alternating,
inequivalent (A and B type) ``chains.''  For zero interchain coupling,
$H^{\prime}=0$ ({\it i.e.} $h\rightarrow \infty$),
the problem is
solved exactly using standard bosonization methods;  we refer to this limit as
the decoupled fixed point, although in reality it is a high-dimensional
manifold of fixed points, parametrized by the various Luttinger exponents of
the
1DEG's.  It is indicated by the point
$\alpha$ in Fig.\ \ref{fig:four}.  By studying the perturbative
renormalization-group (RG) flows in the vicinity of the decoupled fixed
point, as well as the behavior at various anisotropic 2D fixed points, we
determine the qualitative phase diagram and the character of the
excitations in the limit of small, but non-zero interchain coupling.

{\it The decoupled fixed point} Hamiltonian $H^{*}$, consists of a sum
of terms for each decoupled chain.  This problem can be solved
exactly using methods of bosonization\cite{review} to express the Hamiltonian
for
each chain as a sine-Gordon field theory for the spin and charge
degrees of freedom, respectively.  The
electronic field
operators on
each chain can be expressed in terms of the charge and spin
fields, $\phi_{n,s}$ and $\phi_{n,c}$, and their duals,
$\theta_{n,s}$ and $\theta_{n,c}$, as
\ba
\Psi_{n,\sigma}(x)&=& e^{ik_{F}^{n}x} \Psi_{n,\sigma,+}(x)
+  e^{-ik_{F}^{n}x} \Psi_{n,\sigma,-}(x)
\nonumber \\
\Psi_{n,\sigma,\pm}(x)&=& {\cal N}_{n,\sigma} \;
e^{i\sqrt{\pi/2}[\theta_{c}^{n}+\sigma \theta_{s}^{n} \pm
\phi_{c}^{n}\pm\sigma\phi_{s}^{n}]}
\ea
where the Klein factors obey
$\{{\cal N}_{n,\sigma},{\cal N}_{n',\sigma'}\}=\delta_{nn'} \delta_{\sigma
\sigma'}/\pi a$, $a$ is the short-distance cutoff, $\sigma=\pm 1$ for
up and down spins,
$k_{F}^{n}$ is the Fermi wave number of chain $n$, and the bosonic fields
satisfy 
$[\phi_{n,\alpha}(y),\partial_{x}\theta_{m,\alpha^{\prime}}(x)]=i\delta_{n,m}
\delta_{\alpha,\alpha^{\prime}}\delta(x-y)$.  For $n$ even,
the A chains, which  represent the stripe regions where the concentration of
mobile ``doped'' charges is high, are  metallic
but with a spin-gap $\Delta_{s}^{(A)} > 0$, {\it i.\ e.\/} a Luther-Emery
liquid. Microscopically,
one can imagine that each A chain represents the low energy physics of
a doped two-leg or three-leg $t-J$ or Hubbard ladder\cite{EKZ}.  The Fermi
wave-number, $k_{F}^{(A)}$ is therefore far from the commensurate value,
$\pi/2$; experiment\cite{fs} suggests that in a variety of cuprate
superconductors, $k_{F}^{(A)}\approx \pi/4$. The remaining quantities
which characterize the A chains are a charge Luttinger
parameter[\ref{ref:caveat1}],
$K_{c}^{(A)}$, and the charge and spin velocities, $v_{c}^{(A)}$ and 
$v_{s}^{(A)}$. 
The B
chains represent the more lightly doped,
locally antiferromagnetic strips between stripes. We will be 
interested in the case in which
some doped holes have leaked into these strips, so they have no
charge gap, but because they
are still nearly Mott insulating\cite{shulz},
$k_{F}^{(B)}\approx \pi/2$, $K_{c}^{(B)} \approx 1/2$, and
$v_{c}^{(B)}\ll v_{c}^{(A)}$.
We will, however, also consider the cases in which Umklapp scattering
opens a charge gap, $\Delta_{c}^{(B)}$, on chains B, which, in addition,
may or may not have  a spin-gap. In the more interesting
gapless case, spin rotation invariance implies that at low energies the
spin Luttinger exponent, $K_{s}^{(B)} \rightarrow 1$.

{\it The interchain coupling}, $H^{\prime}$, typically\cite{EFKL} generates
interactions that are relevant in the RG sense at the decoupled fixed
point. Starting from a microscopic viewpoint, one would be tempted to
take $H^{\prime}$ to consist of a single particle hopping term which couples
each A chain to its nearest neighbor B chain.  This interaction is
manifestly irrelevant, both
because of the presence of a spin-gap, and because of the mismatch in
Fermi wave numbers, $k_F^{(A)}\ne k_F^{(B)}$.  However, in the initial stages of
renormalization, all imaginable local terms consistent with symmetry
are generated, both relevant and irrelevant.
We will therefore skip this initial step, and
directly study the perturbative $\beta$-functions for potentially
relevant interchain couplings.
Here we assume that $H^{\prime}$ contains only the most 
relevant interactions, between first and second
neighbor chains, for the range of Luttinger exponents discussed above, 
\ba
H^{\prime} & &  =\sum_{n}\int dx \bigg\{
-t_{BB}\sum_{\sigma}[\Psi_{2n-1,\sigma}^{\dagger}\Psi_{2n+1,\sigma}
+ {\rm h. c. }] \nonumber \\
&&- J_{BB}\vec S_{2n-1}\cdot \vec S_{2n+1}-
{\cal J}_{AB}[\hat \Delta_{2n}^{\dagger}\hat \Delta_{2n+1}+ {\rm h. c.}]
\nonumber \\
&&+{\cal J}_{AB}^{\prime}[\hat \Delta_{2n}^{\dagger}(\Psi_{2n-1,\uparrow}
\Psi_{2n+1,\downarrow}+\Psi_{2n+1,\uparrow}
\Psi_{2n-1,\downarrow})+{\rm h. c.}]
\nonumber \\
&& -{\cal J}_{AA}[\hat \Delta_{2n}^{\dagger}\hat \Delta_{2n+2}+ {\rm h. c.}]
\bigg\}
\ea
where $\hat \Delta_{n}\equiv \Psi_{n,\uparrow}\Psi_{n,\downarrow}$
is the singlet pair creation operator, $\vec S$ is the 
spin-density operator, and it is implicitly understood in the above
expression that any piece of the interaction that is rapidly
oscillating (with wave-number $2k_{F}$) is to be omitted.
Here, we have neglected possibly relevant
back-scattering interactions which could potentially promote
charge-density wave formation;  the basis for this is discussed
in Ref. \ref{EFKL}  and below.

The only
non-standard term in $H^{\prime}$ is the term proportional to
${\cal J}_{AB}^{\prime}$ which removes a pair from an A chain, rotates
it by 90$^{o}$, and reinserts it across the two neighboring B chains;
the sign of this term (after renormalization, away from the decoupled, $\alpha$, fixed point)
determines whether the superconducting order is
d-like  or s-like.
Microscopic calculations\cite{2leg} on $t-J$ and Hubbard ladders
lead us to believe that 
most likely  all the pair-tunnelling
terms ${\cal J}_{X} >0$.  
The d-like pairing tendency observed in these calculations
implies ${\cal J}_{AB}^{\prime}>0$, while 
the positivity of the remaining ${\cal J}_{X}$ imply 
an unfrustrated superconducting state;  {\it e.g.} ${\cal J}_{AB}<0$,
would imply ``$\pi$-junctions'' between
neighboring chains.

We begin by considering the regime in which all of the coupling constants in
$H'$ are
small. In this limit, the system is at the decoupled fixed point $\alpha$.  It
is
easy to
determine the (perturbative) role of the various processes in $H'$.
In cases in which a gap prohibits the operation of one of these
interactions in lowest order, that interaction is manifestly
irrelevant, {\it i.\ e.\/} its dimension is infinite.
Otherwise, to leading order in powers of the
interchain couplings, the perturbative $\beta$-functions have the form
\be
\frac{d g}{d \ln a} = (2-D_{g}) g + \ldots
\ee
$D_{g}$ is the scaling dimension of the perturbation with coupling constant
$g$.  For $D_{g} <2$,
the operator is perturbatively relevant, and otherwise it is irrelevant.
The dimension of the various operators 
are listed in Table \ref{table:dimensions}.
Forward scattering interactions
between
the charge currents and densities on neighboring chains are precisely marginal.

We now return to the issue of the CDW couplings between chains; if
relevant, these would lead to an ordered, insulating state.  In Ref. \ref{EFKL},
it was shown that forward scattering interactions between chains, whether
direct or induced by dynamical fluctuations of the stripe geometry, strongly
affect the scaling dimension of these operators, tending to make them less relevant.
In particular, there is a finite regime of parameters, especially when the stripe
fluctuations are significant, where the CDW couplings are irrelevant, and so can
be neglected.

Since in all the cases considered here, the decoupled fixed point is
perturbatively unstable in some way, our next task is to determine where the
RG flows go.

{\it The two fluid fixed point:}  It should be clear from the table
that, under most conditions, all AB couplings are perturbatively
irrelevant at point $\alpha$ of Fig.\ \ref{fig:four}.
Specifically, if we set $K_{c}^{(B)}=1/2$, this is true so
long as $K_{c}^{(A)}< 1$.  (The phase diagrams in Figs.\ 1a, 1c, and 1d
assume this condition is satisfied.)  Even if $K_{c}^{(A)}>1$ (as in
Fig.\ 1b), the AB couplings
are typically more weakly relevant than the couplings between like
chains.  We are therefore led to consider the RG flows in the limit
in which all such couplings are set equal to
zero, so that we have two interpenetrating, and decoupled, but
genuinely two dimensional systems.
This limit is represented by the left-hand edge of the
phase diagrams  in Fig. \ref{fig:four}.

In general, there are many possible higher dimensional fixed points to which
the system could
flow. Because the system can lower its kinetic energy by allowing pairs to 
move between chains\cite{EKZ}, ${\cal J}_{AA}$ is typically relevant-
technically so long as $K_{c}^{(A)}>1/2$.
We will consider exclusively the state in
which the A subsystem has true superconducting long range order and
a full spin-gap. Because $t_{BB}$ always has lower scaling dimension than
$J_{BB}$ so long as $\Delta_{c}^{B}=0$ the most likely situation is
that system B forms a highly anisotropic Fermi liquid.  The
corresponding fixed point is labelled $\beta$ in Figs.\ 1a-1c.
However, either if $\Delta_{c}^{B}>0$, or if the residual interactions
are sufficently strong to drive an instability of the nearly nested
Fermi surface, it is possible that system B will order
antiferromagnetically.  The fixed point with
coexisting superconductivity in A and antiferromagnetism in B is indicated by
$\beta^{\prime}$ in  Fig.\ 1d. (Quantum fluctuations can also lead to a
magnetically disordered insulating state on system B, which
is adiabatically connected to
the state in which $\Delta_{c}^{(B)}$ and $\Delta_{s}^{(B)}$ are non-zero.)

The operators which couple A and B have entirely different
scaling dimensions at the two-fluid fixed points, $\beta$ and $\beta'$, than at
$\alpha$. 
The two-fluid fixed point $\beta$ is unstable, due to the
ordinary proximity effect.  Specifically, in establishing the relevance of
the operators ${\cal J}_{AB}$ and ${\cal
J}_{AB}^{\prime}$, the pair creation operator on A can be replaced 
by their constant expectation
value, while the pairing fields on B operate on a Fermi liquid state.  
For ${\cal J}_{AB}^{\prime}>0$, this superconducting state is  d-wave-like;
the quasi-particle spectrum is:

\begin{table}[b]
\begin{center}
\begin{tabular}{|c||c|c|c|c|}
{} & {$\Delta_{s}^{B}=0$} & {$\Delta_{s}^{B}>0$} &
{$\Delta_{s}^{B}=0$}
& {$\Delta_{s}^{B}>0$}
\\ {} &
{$\Delta_{c}^{B}=0$}  & { $\Delta_{c}^{B}=0$}  &   {$\Delta_{c}^{B}>0$}  &
{$\Delta_{c}^{B}>0$}
\\
\hline
\hline
{$t_{BB}$} & {$(1/4)[K_{c}^{(B)}+$} & { $\infty$} &
{$\infty$} & {$\infty$}
\\
{} & {$1/K_{c}^{(B)} +2]$} & {} & {} & {} \\
\hline
{$J_{BB}$} & {$[K_{c}^{(B)} +1]$} & { $\infty$} & { 1} & { $\infty$ }
\\
\hline
{${\cal J}_{AA}$} & {$1/K_{c}^{(A)}$} & { $1/K_{c}^{(A)}$} & { $1/K_{c}^{(A)}$}
&  {$1/K_{c}^{(A)}$}
\\
\hline
{${\cal J}_{AB}$} & {$(1/2)[1/K_{c}^{(A)}$} &
{$(1/2)[1/K_{c}^{(A)}$} & {$\infty$} & {$\infty$} \\
{} & {$+1/K_{c}^{(B)}+1]$} & {$+1/K_{c}^{(B)}]$} & {} & {} \\
\hline
{${\cal J}_{AB}^{\prime}$} & {$(1/4)[2+2/K_{c}^{(A)}$} & { $\infty$ } & {
$\infty$ } &
{ $\infty$ } \\
{} & {$ +K_{c}^{(B)}+1/K_{c}^{(B)}]$ } & {} & {} & {}\\
\end{tabular}
\caption{ Scaling dimensions of the interchain couplings at the
decoupled fixed point with $\Delta_s^A>0$ and $\Delta_c^A=0$.}
\label{table:dimensions}
\end{center}
\end{table}

\ba
E^{2}(\vec k)&&
=[v_{F}(k_{x}-k_{F}^{B}) - 2t_{BB}\cos(k_{y}L)]^{2}  \\
&&+ <\hat \Delta_{A}>^{2}[{\cal
J}_{AB}-2{\cal
J}_{AB}^{\prime}\cos(k_{y}L)]^{2}\nonumber
\ea
\noindent 
where x and y refer, respectively to the directions parallel 
and perpendicular to the chains, $v_{F}$, $t_{BB}$,  ${\cal J}_{AB}$
and ${\cal J}_{AB}^{\prime}$ are now to be interpreted as renormalized 
parameters at 
$\beta$, and $L$ is the spacing between chains.
There remains the quantitative issue, which depends on the microscopic
details, of the relative magnitudes of these couplings.
If $|{\cal J}_{AB}|<2|{\cal
J}_{AB}^{\prime}|$, the quasi-particle spectrum is gapless.  This is
a stable phase of matter, the nodal superconductor, described by
the fixed point $\gamma$ in Fig.\ 1.  We have
assumed that this inequality is satisfied in Figs.\ 1a and 1b, where
the RG flows run directly from $\beta$ to
$\gamma$.  However, if $|{\cal J}_{AB}|> 2|{\cal
J}_{AB}^{\prime}|$, the flows run from $\beta$ to the fully gapped
superconducting state, signified by the stable fixed point $\delta$ in
the figure, as in Fig.\ 1c.

By contrast, at the fixed point $\beta^{\prime}$,
there are no finite dimensional operators which couple the two fluids
since the antiferromagnet has a charge gap and the
superconductor has a spin-gap. This fixed point describes a stable
phase with two coexisting order parameters, and a complete
gap in the quasi-particle spectrum. Evidence of a phase with
coexisting magnetic and superconducting order, presumably rendered glassy 
by quenched disorder, has been presented \cite{lnsco,lco}; 
it is a prediction of the present study that the ordered state should 
have a fully gapped quasi-particle spectrum.

{\it Fully coupled fixed points:}  In addition to the stable fixed
points $\gamma$ and $\delta$, described above, there are a number of
other unstable fixed points whose existence is dictated by the
topology of the phase diagram (assuming that the transitions are second order).

The phase
transition between the nodal and the gapped superconductor
is governed by the fixed
point $C_{1}$ in the figure.  The universal properties of this
transition can be studied\cite{sachdev} in the weakly interacting limit.  
This transition, as approached from the nodal phase, is triggered by
deforming the gap function (or band structure) such that two nodal
points approach each other, and at the critical point, coalesce.
As a consequence, at the single nodal point, the quasi-particle
velocity in one direction vanishes.  Beyond the critical point, a gap
opens in the spectrum.  By naive power counting, four fermion
interactions are irrelevant at this fixed point, so this is all there
is to it!
Even when ${\cal J}_{AB}$ is perturbatively irrelevant at $\alpha$,
strong enough pair-tunnelling will certainly produce
superconductivity in B by the proximity effect.  Moreover, the
resulting state will be the fully gapped, ordered superconductor.
Thus, there must exist a highly unstable critical point, $C_{2}$, 
similar to the smectic metal to superconductor
critical point of Ref.\ \cite{EFKL}.

The critical point $C_{3}$ in Fig. 1d is similar to one
mentioned by Vojta {\it et.\ al.\/} \cite{sachdev},  
for a transition, within
a nodal superconducting phase, to a state with broken translational
symmetry with an ordering vector which at $C_3$ spans the nodal points. 
Conversely, the robustness of the
nodal superconducting phase embodied in Figs.\ 1a, 1b, and 1d,
mirrors the asymptotic decoupling  of the nodal
quasi-particles from fluctuations associated with such a transition
when the ordering vector does not span the nodes.
The (as yet not analyzed) multi-critical point $C_{4}$
is required by a minimal consistent construction of the RG flows.

Finally, it is clear that at the fixed
point $\gamma$ the nodal quasi-particles are well defined elementary
excitations of the system at arbitrarily low energy and long
wave-length. This is consistent with a widely held belief that such
excitations must exist, even though the high temperature
superconductors are far from the BCS limit. However, what is also
clear from the circuitous flows that lead to this fixed point in
Figs. 1a and 1b is that these quasi-particles can, under appropriate
circumstances, be much less robust than the superconducting state,
itself. Thus, even well below
$T_{c}$, where superconducting order is well established, the nodal
quasi-particles can still be ill-defined objects, and only become
sharp at very low temperatures.  It is possible that this observation
reconciles the strong evidence\cite{taill} from thermal conductivity
of well defined
nodal quasi-particles at  
$T \ll T_{c}$ with
the  evidence from photo-emission\cite{valla} of their non-existence
down to temperatures of order $T_{c}/2$.

We thank D.\ Orgad and S.\ Sachdev for useful discussions.
EF, SAK and VO
were participants in the High $T_c$ Program at ITP-UCSB.
This work was supported in part by the NSF grants DMR98-08685
at UCLA (SAK and VO),
NSF DMR98-17941 at UIUC (EF), NSF PHY99-07949 at ITP (EF,SK,VO)
DOE grants No. 
DE-AC02-76CH00016 at BNL and DE-FG03-00ER45798 at UCLA,
and 
a STINT (Sweden) fellowship (MG).


\end{multicols}

\begin{references}

\bibitem{schrieffer}
J.\ R.\ Schrieffer, {\sl The Theory of
Superconductivity}, Addison Wesley (Redwood City, 1964).

\bibitem{carlson}
See,  E.\ W.\ Carlson {\it et.\ al.\/},
Phys.\ Rev.\ B {\bf 62}, 3422, 2000,
and references therein.

\bibitem{PNAS}
V.\ J.\ Emery {\it et al},
Proc.\ Natl.\ Acad.\ Sci.\ 
 {\bf 96}, 8814 (1999).

\bibitem{fedorov}
A.\ V.\ Fedorov, \prl\ {\bf 82}, 2179 (1999);
D.\ L.\ Feng {\it et.\ al.\/}, Science {\bf 289}, 277 (2000);
H.\ Ding  {\it et al} cond-mat/0006143.

\bibitem{fs}
X.\ J.\ Zhou {\it et.\ al.\/}, Science {\bf 286}, 5438 (1999);
X.\ J.\ Zhou {\it et.\ al.\/}, cond-mat/0009002.

\bibitem{2leg}
S.\ White and D.\ J.\ Scalapino, \prb {\bf 57}, 3031 (1998);
T.\ M.\ Rice {\it et.\ al.\/}, \prb {\bf 56}, 14655 (1997).

\bibitem{shulzd}
H.\ Schulz, \prb {\bf 53}, R2959 (1996).

\bibitem{review}
See V.\ J.\ Emery in
{\it Highly Conducting 1D Solids}, eds.\ J.\ T.\ Devreese,
{\it et al} (Plenum, New York, 1979).


\bibitem{EKZ}
V.\ J.\ Emery, S.\ A.\ Kivelson and O.\ Zachar,
\prb {\bf 59}, 15641 (1999).

\bibitem{caveat1}
For a single chain with repulsive interactions $K_c^A<1$. However,
for a multi-chain case
what matters is the effective $K_c^A$ which can be
renormalized to values greater or smaller than $1$, but typically close to $1$.
\label{ref:caveat1}

\bibitem{shulz}
H.\ Schulz, \prb {\bf 22}, 5274 (1980).

\bibitem{EFKL}
S.\ A.\ Kivelson, E.\ Fradkin and V.\ J.\ Emery, Nature {\bf 393}, 550
(1998);
V.\ J. Emery {\it et.\ al.\/}, \prl {\bf 85}, 2160 (2000).
\label{EFKL}

\bibitem{lnsco}
J.\ Tranquada {\it et.\ al.\/}, Nature {\bf 375}, 561
(1995).

\bibitem{lco}
K.\ Yamada, {\it et.\ al.\/}, {\prl} {\bf 75}, 1626-1629 (1995);
Y.\ S.\ Lee {\it et.\ al.\/}, {\prb} {\bf 60}, 3643 (1999).

\bibitem{sachdev}
M.\ Vojta, Y.\ Zhang and S.\ Sachdev, Phys.\ Rev.\ Lett.\ {\bf 85}, 4940 (2000)

\bibitem{taill}
See M.\ Chiao {\it et.\ al.\/}
\prb {\bf 62}, 3554 (2000).

\bibitem{valla}
T.\ Valla {\it et.\ al.\/},
Science {\bf 285}, 2110 (1999).
\label{ref:valla1}

\end{references}
\end{document}